\documentclass[twocolumn,showpacs,preprintnumbers,amsmath,amssymb]{revtex4}
\usepackage{graphicx}

\begin{document}

\title{An experimentally testable proof of the discreteness of time}
\author{Guang Ping He}
\email{hegp@mail.sysu.edu.cn} \affiliation{School of Physics \&
Engineering and Advanced Research Center, Sun Yat-sen University,
Guangzhou 510275, China}

\begin{abstract}
By proposing a paradox between the impossibility of superluminal
signal transfer and the normalization condition of wavefunctions, we
predict that when a change happens to the conditions that
determining the status of a quantum system, the system will show no
response to this change at all, until after a certain time interval.
Otherwise either special relativity or quantum mechanics will be
violated. As a consequence, no physical process can actually happen
within Planck time. Therefore time is discrete, with Planck time
being the smallest unit. More intriguingly, systems with a larger
size and a slower speed will have a larger unit of time. Unlike many
other interpretations of the discreteness of time, our proof can be
tested, at less partly, by experiments. Our result also sets a limit
on the speed of computers, and gives instruction to the search of
quantum gravity theories.
\end{abstract}


\maketitle

\section{Introduction}

How does time flow? This mystery puzzled people for centuries. In
the age of Newton's classical mechanics, time was regarded as a
continuous coordinate that flows independently of any other object
and event. With the advance of modern physics, people started to
aware that time cannot exist solely without involving any physical
process. That is, time can only be sensed and measured when changes
occur to the status of objects. As indicated by quantum mechanics,
there should be a smallest unit of time called Planck time, which is
the limit of time that can be measured due to the uncertainty
principle. Thus physics cannot reason in a meaningful way what
happens within a time interval shorter than Planck time. Here we
show that it is not only the limit of measurement. Instead, we give
a simple proof that no physical process can actually happen within
Planck time, otherwise either special relativity or quantum
mechanics will be violated. Therefore it is proven that time is
discrete instead of continuous.

\section{A paradox between special relativity and quantum mechanics}

According to the theory of special relativity, no signal can travel
faster than the speed of light. On the other hand, quantum mechanics
claims that any physical system is completely described by a
wavefunction, which has to be normalized. Now let us show that there
is a paradox between the two. Suppose that we want to induce a
change on the wavefunction of a quantum system. Then we need to make
a change on the elements which determine the wavefunction of the
system, e.g., the potential, the status of the boundary, etc.. How
fast will the wavefunction show a change after these elements
changed? For simplicity, let us consider the state of a particle in
a one-dimensional finite square well with a potential $V_{0}$ as
shown in Fig.
1(a). Denote the normalized wavefunction of the particle in this case as $%
\psi _{0}$. At time $t_{1}$, the potential at point $A$ suddenly changes to $%
V_{1}$, as illustrated in Fig. 1(b). Suppose that $\psi _{1}$ is the
normalized wavefunction satisfying the current value of the
potential. Then what is the minimal time for the state to change
from $\psi _{0}$ to $\psi _{1}$?

It is important to note that even though such kinds of questions
seem quite usual in quantum mechanics, in literature they were
always solved under nonrelativistic approximation, despite that this
was not clearly stated most of the time (see e.g., Refs.
\cite{ft35,ft34}). That is, it is assumed that
at any given time $t$, the wavefunction $\psi (t)$ satisfies the Schr\"{o}%
dinger equation of the same $t$. This actually means that the change
to the wavefunction occurs in the whole space instantaneously when
the Hamiltonian changes. But this will violate special relativity as
shown below. Suppose
that two people Alice (located at point $A$) and Charlie (located at point $%
C $, where the distance between points $A$ and $C$ is $l$) want to
communicate. They prepared $N$ ($N$ is sufficiently large) copies of
the system shown in Fig. 1(a) beforehand. At time $t_{1}$, if Alice
wants to send the bit $1$, she makes the potential $V_{0}$ of the
first $N/2$ systems at point $A$ change to $V_{1}$ simultaneously,
while leaving the last $N/2$
systems unchanged. Else if she wants to send the bit $0$, she keeps all the $%
N$ systems unchanged. At time $t_{1}+\Delta t$ Charlie measures all
the $N$ systems at point $C$. If the probability of finding the
particle at point $C$ in the first $N/2$ systems can be considered
equal to that of the last $N/2$ systems within the variation range
allowed by statistical fluctuation, he assumes that the bit sent by
Alice is $0$. Else if the probabilities of finding the particle at
point $C$ look much different in the two halves of the systems, he
assumes that the bit is $1$. With this method, a
superluminal signal can be transferred from point $A$ to $C$ if $l>c\Delta t$%
, where $c$ denotes the speed of light. Thus special relativity is
violated if we assume that the wavefunction can change from $\psi
_{0}$ to $\psi _{1}$ in the whole space instantaneously at time
$t_{1}$.

Therefore, it seems natural to assume that the response of the
wavefunction corresponding to the change of the potential at point
$A$ should propagate along the $x$ axis with a finite speed $v$
($0<v\leq c$). However, this will cause trouble to the normalization
of the wavefunction. Suppose that at time $t_{1}$ the potential at
point $A$ changes, and at time $t_{1}+l/v$ the change of the
wavefunction from $\psi _{0}$ to $\psi _{1}$\ occurs to the
locations between points $A$\ and $C$, while the wavefunction at the
right side of $C$ has to remain strictly unchanged due to the
impossibility of superluminal signal transfer. Such a wavefunction
is plotted as the solid blue curve in Fig. 1(c). Then we can
immediately see from the figure that the resultant wavefunction at
this moment is no longer normalized.

There is also other possible way of evolution of the wavefunction
which can keep the normalization condition unbroken. However, as
shown in the appendix, such a solution will not satisfy the basic
equations of quantum mechanics, i.e., quantum theory will be
violated too.

Thus we found a paradox between the impossibility of superluminal
signal transfer and the normalization condition of wavefunction.
Though it was well-known that the theories of relativity and quantum
mechanics do not go well with each other, the current paradox
reveals yet another conflict between the theories which does not
seem to have been reported before. Also, it does not involve the
transformation between reference frames, so it cannot be solved
simply by replacing Schr\"{o}dinger equation with Klein-Gordon or
Dirac equation. Therefore it differs by nature from previously known
conflicts between the two theories, and put forward a new challenge
to our understanding on the quantum world.

\section{The discreteness of time}

Intriguingly, we find that this paradox can be solved if we adopt
the bizarre idea that time is discrete. As shown above, relativity
does not allow the change of wavefunction to occur before the time
$t_{1}+l/v$ for any location whose distance from point $A$ is larger
$l$. Meanwhile, quantum mechanics does not allow the wavefunction to
change part by part. As a consequence, to obey both theories
simultaneously, logically the wavefunction has to evolve in the
following way. After the potential at point $A$ changed at time
$t_{1}$, the wavefunction should show no response at all during a
period of time $\tau $. Then at time $t_{1}+\tau $ or some point
later, the wavefunction in the whole space changes simultaneously to
keep the normalization condition unbroken. Here%
\begin{equation}
\tau =L^{\prime }/v\geq L/v,  \label{eq1}
\end{equation}%
where $L$ is the distance between points $A$ and $B$, which can be
regarded as the effective size of the system. $L^{\prime }$ is the
distance between points $A$ and $B^{\prime }$ or points $A$ and
$A^{\prime }$, depending on which one is larger. The location of
points $A^{\prime }$ and $B^{\prime }$ are determined by the
wavefunctions, in such a way that the difference between $\psi _{0}$
and $\psi _{1}$ at the left side of $A^{\prime }$ and the right side
of $B^{\prime }$ is completely drowned by statistical fluctuation,
so that it will not lead to any detectable superluminal signal from
point $A$ to these regions when the wavefunction changes from $\psi
_{0} $ to $\psi _{1}$.

In the spirit of Newton's first law, any physical system will
persist in its state of motion unless being applied with an
inducement. Meanwhile, since quantum mechanics is recognized as the
complete description of the physical world, any physical process can
be viewed as the change of the wavefunction
of the system under a certain inducement on a certain point. Therefore Eq. (%
\ref{eq1}) sets a limit on how fast any physical process can occur.
That is, when any inducement is applied on a system with size $L$,
no change can happen to the state of the system within the time
$\tau $. This conclusion covers all systems including any object we
want to measure, as well as all apparatus we use as timekeepers or
detectors to measure other objects. Now consider the lower bound of
$\tau $ for any system. Due to the uncertainty principle of quantum
mechanics, the minimal size of any physical system that can be
reasoned in a meaningful way is Planck length $l_{P}\simeq
1.616\times 10^{-35}meter$. Meanwhile, the theory of relativity requires $%
v\leq c$. Thus we have%
\begin{equation}
\tau _{\min }\geq l_{P}/c\equiv t_{P}\simeq 5.39\times 10^{-44}\sec
. \label{eq2}
\end{equation}%
Here $t_{P}$ is known as Planck time, which was already recognized
as the minimum of time that can be measured. Our result suggests
that the significance of $t_{P}$ is more than that. Any physical
change can only happen after a time which is not less than $t_{P}$.
Within a time interval of $t_{P}$, any physical system simply
persists in its previous state. Therefore, according to the modern
understanding of time, nothing happens within $t_{P}$ so that there
is no further division of time possible in this range. In this
sense, time is discrete, with $t_{P}$ being the minimal unit.
Because the value of $t_{P}$ is so small, it is not surprising that
the discreteness of time is less noticeable in practice, and
previous nonrelativistic treatment of quantum mechanical problems
\cite{ft35,ft34} seems fine in most cases.

Note that even if the minimal size limit $l_{P}$ could be somehow
broken in the future, it is still natural to believe that as long as
a system exists physically, its size has to be a finite
non-vanishing value. Therefore according to Eq. (\ref{eq1}), time
still cannot be made continuous, though the exact value of the
minimal unit might differs from $t_{P}$.

The above analysis is based on the assumption that the change of the
potential from $V_{0}$ to $V_{1}$\ is completed instantly. Some may
wonder how this can be possible if $t_{P}$ is the minimal unit of
time. Also, it would be interesting to ask what will happen if the
potential changes more than once within $t_{P}$. We believe that
these problems should be understood as follows. Even if there exists
a change of the potential (or any other inducement) which could be
so fast that it occurred and completed instantly, our result above
showed that the response of any system to this change cannot occur
within $t_{P}$. Therefore the response will surely take more time to
occur if the change takes a finite time to complete, so that it will
not conflict with the conclusion that no physical process can occur
within a time interval less than $t_{P}$. Also, any change of the
potential
has to be made by a certain physical apparatus, which is also limited by $%
t_{P}$. Once the potential has a change at time $t_{1}$, no physical
process can change it again before time $t_{1}+t_{P}$. That is, the
change should also be considered as discrete instead of continuous.
Therefore there does
not exist the case where the system encounters a series of changes within $%
t_{P}$.

\section{Experimental test}

Now back to the system with a size $L>>t_{P}$. It is interesting to
notice
from Eq. (\ref{eq1}) that the system has its own minimal unit of time $\tau $%
. The larger the size $L$ is, the slower the system can evolve. Note
that for complicated systems containing more than one particle, $L$
should be understood as the minimal localization length of the
particles in the system, rather than the overall size. Therefore
macroscopic systems, e.g., human bodies or planets, do not mean
having a tremendous $\tau $, because they contain plenty of
particles which are highly localized on the microscopic scale. But
for a system which is relatively large while having a simple
structure, if we can keep all the particles on extended states whose
localization length is comparable with the size of the system, then
it may become possible to observe a larger discreteness of time.
Therefore, though we cannot test directly our above interpretation
of the discreteness of time with systems having the size of Planck
length because we cannot find detectors smaller then they do, we can
perform indirect experimental test with larger systems.

For example, we can stimulate the device in Fig. 1 with cold atoms
or quantum dots. We also use a detector to measure whether the
particle inside the well can be found within a fixed region inside
the potential well. This region serves as point $C$ in Fig. 1(c). To
test whether time is discrete, first we repeat this experiment many
times to get an estimation of the probability $p_{0}$ of finding the
particle around point $C$ when the potential is set to $V_{0}$.
Secondly, we re-initialize the system, i.e., prepare such a system
again with the potential $V_{0}$ and keep it unmeasured. Then change
the potential at point $A$ from $V_{0}$ to $V_{1}$. After the change
we wait for a time interval $t_{x}<l/c$ where $l$ is the distance
between points $C$ and $A$. Now we measure whether the particle can
be found at point $C$. Repeat this for many times too, so we can get
an
estimation of the probability $p_{x}$ of finding the particle around point $%
C $ at time $t_{x}$ after the potential changed. Third, we
re-initial the system again. Change the potential from $V_{0}$ to
$V_{1}$, and wait for a time interval $t_{y}$ ($l/c<t_{y}<L/c$).
Here $L$ is the width of the well. Then we measure whether the
particle can be found at point $C$. Repeat this also for many times
and we can get an estimation of the probability $p_{y}$ of finding
the particle around point $C$ at time $t_{y}$ after the potential
changed. Finally we compare whether $p_{0}$, $p_{x}$ and $p_{y}$ are
equal within the variation of statistical fluctuation. Then we can
have the following conclusion.

(1) If $p_{0}\neq p_{x}$, then it seems to enable superluminal
signal transfer and thus violates the theory of relativity.

(2) Else if $p_{0}=p_{x}\neq p_{y}$, then the theory of relativity
is obeyed but quantum mechanics seems to be violated.

(3) Else if $p_{0}=p_{x}=p_{y}$, then it proves that our above
interpretation of the discreteness of time is correct.

For a more rigorous test on the validity of the normalization
condition, we can keep changing the location of the detector (i.e.,
point $C$), and measure the probabilities of finding the particle at
different positions at a given time after the potential changed from
$V_{0}$ to $V_{1}$. If for a period of time after the potential
changes, the measured value of the probability at every position
remains unchanged, and at a later time we find that all in a sudden,
the probability at every point shows difference from its previous
value, then we can conclude that the wavefunction in the whole space
indeed changes simultaneously. Else if we find that at a given time,
the probabilities change at some positions while remain unchanged at
the others, then we can deduce the form of the current wavefunction
and check whether the normalization condition could be broken.

In these experiments, we're mostly interested on whether the
probabilities measured at different time or positions are equal or
not. The exact values of the probabilities are not very important as
long as we do not need to check the normalization condition in exact
numbers. Therefore it does not matter how much the detection
efficiency of the detector is. As long as the efficiency remains
stable during the experiments, the result will be valid.

But we have to notice that both $t_{x}$ and $t_{y}$ are very small
time intervals. For instance, the size $L$ of quantum dots are
usually hundreds of nanometers, so $L/c$ is at the order of
magnitude of femtoseconds. Thus it will be hard to measure $t_{x}$
and $t_{y}$ precisely with current technology. Nevertheless, Eq.
(\ref{eq1}) shows that the minimal time interval $\tau $ of a system
is determined by $L/v$ instead of $L/c$. If $v$ is significantly
smaller than $c$, then we can expect a much longer time interval
$t_{z}>L/c>t_{y}$, during which the system still does not evolve as
long as $t_{z}<L/v$. Since the exact value of $v$ is unknown to us
so far, in experiments we can measure the probability $p_{s}$ of
finding the particle around point $C$ at time $t_{s}$ after the
potential changes, where $t_{s}$ is the shortest time interval we
can achieve with current technology. If we find $p_{s}=p_{0}$ within
the variation of statistical fluctuation, then it proves our above
interpretation of the discreteness of time, while also indicates
that $v<L^{\prime }/t_{s}\simeq L/t_{s}$. By
increasing $t_{s}$ gradually and repeating the experiment until we find $%
p_{s}\neq p_{0}$, the speed $v$ can be more rigorously determined.

Here we would like to discuss the speed $v$ a little further. $v$
describes how fast a system responses to the change of the status of
the boundary at one side, therefore its value is related with the
understanding on how the particle in the system \textquotedblleft
knows\textquotedblright\ the status of the boundary. This is a
question of quantum interpretation theory which is beyond the
standard framework of quantum mechanics. Indeed, previous quantum
mechanical calculations all settled with the nonrelativistic
approximation (e.g., Refs. \cite{ft35,ft34}) where $v$ is treated as
infinite. Therefore these theories are insufficient for calculating
the value of $v$ without the help of quantum interpretation. In the
square well problem we considered here, however, different quantum
interpretation theories may have different understanding on how the
particle \textquotedblleft knows\textquotedblright\ the status of
the boundary of the well, so they may predict different values of
$v$. Thus it will be very useful if we could measure the speed $v$
experimentally. While it may not be easy to implement the above
experimental proposal to measure $v$ precisely
with current technology, a very feasible scheme was proposed in Ref. \cite%
{He}, which can measure a similar speed in double-slit interference
using state-of-the-art technology. Though that experiment has not
been carried out yet, it was also shown in the same reference that
the speed it measures may very probably equal to the classical speed
$v_{0}$ of the particle in the system if the mainstream quantum
interpretation theories are correct. This is because in double-slit
interference, the mainstream interpretations believe that the
particle \textquotedblleft knows\textquotedblright\ the status of
the slits by reaching them by itself. Therefore, it is reasonable
to believe that the speed $v$ in our current case also equals to the speed $%
v_{0}$ of the particle. If so, we can expect a very low $v$ from
systems at low energy levels. Then it will be easier to observe a
larger $\tau $ in the above experiments. Moreover, $v=v_{0}$ can
also explain why the system needs to wait a time interval $\tau $
before its wavefunction starts to evolve. As we know, the
wavefunction (e.g., $\psi _{1}$ in Fig. 1(b)) is not determined
merely by the status of the potential at one side of the well.
Instead, the potential on anywhere of the whole system has its share
of contribution. Therefore, the particle needs time $\tau \geq
L/v_{0}$ to travel from one side of the well to the other, so that
it \textquotedblleft knows\textquotedblright\ the status of the
potential and the boundary at every point of the system before it
\textquotedblleft decides\textquotedblright\ how the wavefunction
should evolve to. Of course, all quantum interpretation theories are
still in need of more experimental support at the present moment.
Therefore it is still too early to reach any conclusion
theoretically, before the above experimental proposals are
implemented and prove whether $v=v_{0}$ or not.

\section{Applications}

As a corollary of the discreteness of time, there exists a limit for
the speed of all kinds of computers, either classical or quantum
ones. Since the state of a register of the computer cannot evolve
within $\tau $, every step of the instruction on the register needs
at least a time interval of $\tau $
to complete. Thus the maximum operational speed on a single register is $%
1/\tau $ $IPS$ (Instructions per second). If $t_{P}$ is indeed the
minimal unit of time, then the maximum speed is $1/t_{P}\simeq
1.86\times 10^{43}IPS$ per register. Of course, a computer can
contain many registers that operate in parallel. Therefore the total
speed will rise with the increase of the number of the registers.

We should note that the above analysis is based on the assumption
that both special relativity and quantum mechanics are valid on any
scale. This assumption seems to be valid on most scales, even down
to a few nanometers. Therefore our above analysis and experimental
proposal are valid for systems with a larger size $L$, e.g., quantum
dots and cold atoms. But currently there is no proof that special
relativity and quantum mechanics must remain valid in the range of
Planck length and Planck time, despite that neither counterexample
was found so far. If either of the theories fails on this scale,
then there may be physical processes happening within the range of
Planck time. Ironically, if the above interpretation on the
discreteness of time is truth, then as we mentioned, all timekeepers
and detectors are bounded by the limit of the discreteness too. Thus
we cannot perform experimental test directly on this scale despite
that we can perform indirect test on larger scales. Nevertheless, it
does not mean that our above analysis is futile even on the scale
where special relativity or quantum mechanics becomes invalid.
Currently there are many attempts to develop new theories (e.g.,
loop quantum gravity theory) trying to describe the events within
the scale of Planck time, e.g., the first few moment of our universe
just after it was born from the Big Bang. Our current result
indicates that if there indeed exists a theory capable of handling
the physical processes in any small time interval, i.e., time is
treated as continuous, then it should be better
\textit{incompatible} with either the impossibility of superluminal
signal transfer or the normalization condition of wavefunction (or
even both). Or it should find an even smaller unit as a replacement
for Planck time to describe the discreteness of time. Otherwise it
will have a hard time solving the paradox between relativity and
quantum mechanics we proposed above. This is in agreement with a
recent proposal \cite{ft53}, which prefers an indefinite causal
structure of the theory.

We thank Prof. Sofia Wechsler for valuable discussions. The work was
supported in part by the NSF of China under grant Nos. 10975198 and
10605041, the NSF of Guangdong province under grant No.
9151027501000043, and the Foundation of Zhongshan University
Advanced Research Center.

\appendix

\section{Another solution that will violate quantum mechanics}

Alternatively, we find that there exists another solution to the
above paradox between the impossibility of superluminal signal
transfer and the normalization condition of wavefunctions, which
does not require the
discreteness of time. That is, when the potential of the well changed from $%
V_{0}$ to $V_{1}$ at time $t_{1}$, the wavefunction evolves from
$\psi _{0}$ to $\psi _{1}$ in the following way. At time
$t_{1}+\delta t$, the wavefunction $\psi _{\delta }$ varies from
$\psi _{0}$ with a wave-like shape in a small region around point
$A$ only. At one side of point $A$ the wavefunction rises a little,
while at the other side of point $A$ it drops the same amount, so
that the overall wavefunction is still normalized. An example of
such a wavefunction is illustrated in Fig. 2. The width $d$ of the
varied region at each side of point $A$ grows as $\delta t$\
increases, but for any given $\delta t$, it always satisfies $d\leq
c\delta t$ so that no superluminal signaling occurs. Thus the above
paradox is avoided.

Nevertheless, we must notice that even though such an evolution of
the wavefunction does not violate the normalization condition, it
may still violate quantum mechanics since $\psi _{\delta }$ is not
the solution to Schrodinger (or Klein-Gordon/Dirac) equation with a
Hamiltonian corresponding to $V_{0}$ or $V_{1}$. That is, standard
quantum mechanical formulas alone are insufficient to describe the
behavior of such a wavefunction. We will have to find new formulas
and even new postulations to explain why the system should take such
a wavefunction, how it will finally evolve to $\psi _{1}$, and what
determines the shape of $\psi _{\delta }$
(e.g., the phase, amplitude, and speed of the wave-like variation between $%
\psi _{\delta }$ from $\psi _{0}$), which seem to exceed the
framework of standard quantum mechanics. Therefore, the existence of
such a solution does not conflict with the conclusion that the
discreteness of time should be the solution if we want to keep both
special relativity and quantum mechanics unbroken.

\begin{figure}[tbp]
\includegraphics{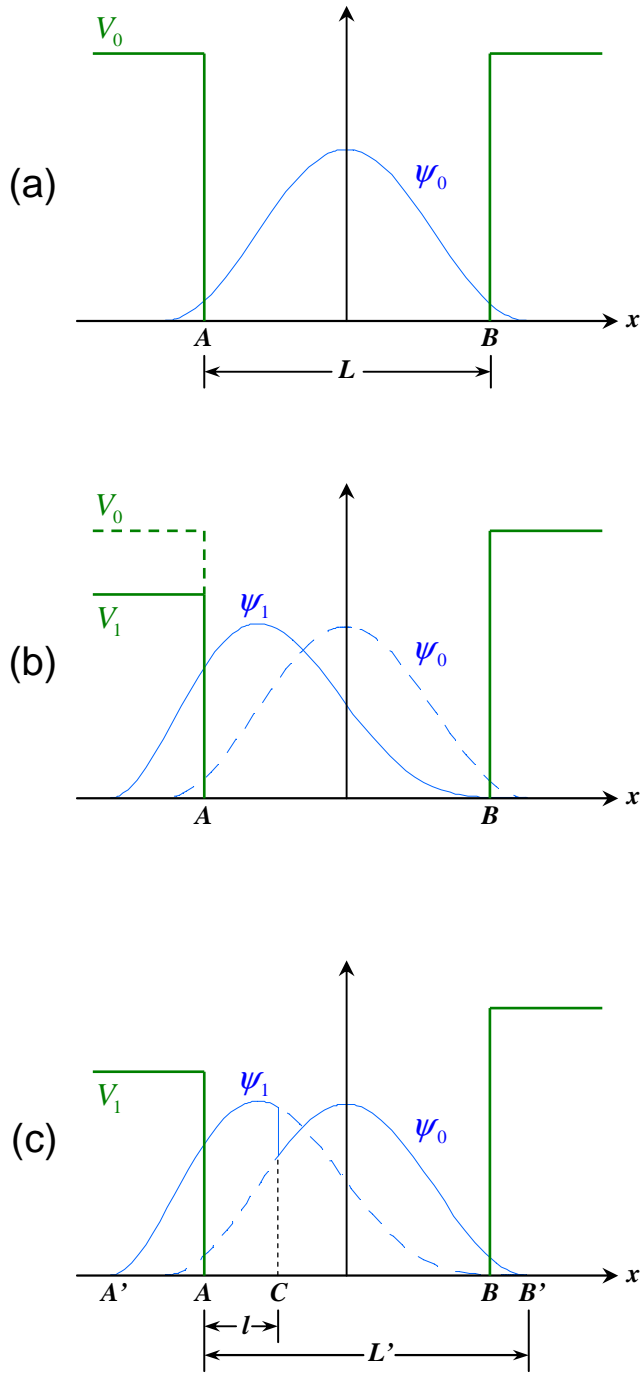}
\caption{The wavefunctions in a square well. (a) The wavefunction $\protect%
\psi _{0}$ when the potential is $V_{0}$. (b) The wavefunction $\protect\psi %
_{1}$ (solid blue curve) when the potential is $V_{1}$ (solid green
curve).
(c) The wavefunction at time $t_{1}+l/v$ (solid blue curve) as a mix of $%
\protect\psi _{0}$ and $\protect\psi _{1}$.} \label{fig:epsart}
\end{figure}

\begin{figure}[tbp]
\includegraphics{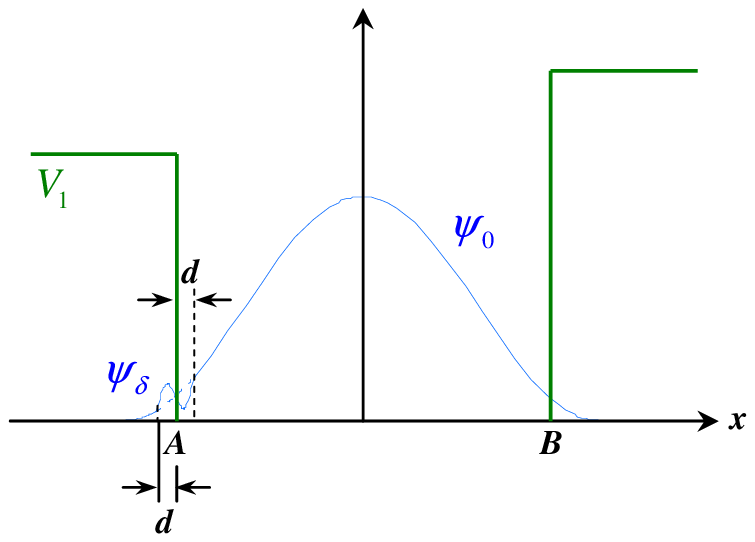}
\caption{An example of the variation between the wavefunction $\psi
_{\delta }$
(solid blue curve) and $\psi _{0}$\ (dashed blue curve) at time $%
t_{1}+\delta t$.} \label{fig:epsart}
\end{figure}


\begin{thebibliography}{9}
\bibitem{ft35} M. Moshinsky, \textit{Phys. Rev.} \textbf{81}, 347 (1951).

\bibitem{ft34} A. del Campo, J. G. Muga, and M. Kleber, \textit{Phys. Rev.}
A \textbf{77}, 013608 (2008).

\bibitem{He} G. P. He, arXiv:0907.1974.

\bibitem{ft53} L. Hardy, arXiv:0804.0054. To appear in A. Bokulich and G.
Jaeger (Eds.), \textit{Philosophy of Quantum Information and Entanglement},
Cambridge University Press.
\end{thebibliography}
\end{document}